\documentclass[a4paper]{jpconf}
\usepackage{graphicx}
\usepackage{url}
\usepackage{amssymb}
\usepackage{amsmath}
\usepackage{epsfig}
\usepackage{amsfonts}
\usepackage{graphicx}

\input{tcilatex}

\begin{document}

\title{Resonances $f_{0}(1370),$ $f_{0}(1500)$ and $f_{0}(1710)$ within the
extended Linear Sigma Model}
\author{Stanislaus Janowski and Francesco Giacosa}

\address{Institute for Theoretical Physics, Goethe-University, Max-von-Laue-Str. 1, 
\\60438 Frankfurt am Main, Germany}

\begin{abstract}
In the framework of the extended Linear Sigma Model (eLSM) we investigate
the masses and decays of the three scalar-isoscalar resonances $f_{0}(1370)$%
, $f_{0}(1500)$ and $f_{0}(1710)$. The degrees of freedom of the eLSM are
(pseudo)scalars and (axial)vectors as well as the scalar glueball. Although
still preliminary, present results, based on the physical masses of the
above mentioned resonances, show that $f_{0}(1710)$ is the predominantly
glueball state. However, acceptable decays for this resonance can be
obtained only for a (very) large gluon condensate.
\end{abstract}

\section{Introduction}

In this work we study the scalar-isoscalar resonances $f_{0}(1370)$, $%
f_{0}(1500)$ and $f_{0}(1710)$ \cite{PDG} in the framework of the extended
Linear Sigma Model (eLSM) \cite{susden, dilaton, dick, dilnf3}. The eLSM is
built accordingly to two fundamental properties of QCD: chiral symmetry and
dilatation invariance, the former being spontaneously broken by a
Mexican-hat potential, the latter being explicitly broken in such a way to
mimic the trace anomaly of QCD \cite{schechter, migdal}. As a consequence of
these two requirements, the eLSM Lagrangian contains only a finite number of
terms. Moreover, (axial)vector d.o.f. are included from the very beginning
in the effective model. The inclusion of the latter in a chiral invariant
way makes the model more complete and has a strong influence on the phenomenology.

In Ref. \cite{susden} the eLSM has been first studied, both in the baryonic
and mesonic sectors, in presence of two-flavours, $N_{f}=2.$ The inclusion
of the dilaton/glueball field for $N_{f}=2$ has been carried out in\ Ref. 
\cite{dilaton}, where it has been shown that the glueball is predominantly
contained in either $f_{0}(1500)$ or $f_{0}(1710)$ (the former case being
favoured). Still, the study was not complete because scalar mesons
containing the $s$-quark were not included. A detailed study of the eLSM for
three flavours, $N_{f}=3$, has been performed in\ Ref. \cite{dick}, where a
very good description of various meson masses and decays was achieved. Yet,
the dilaton, although formally included to justify the form of the
Lagrangian, was inert (i.e., with zero width, in agreement with large-$%
N_{c}$ limit). The scalar-isoscalar system was studied with the nonstrange
quarkonium field $\sigma _{N}\equiv (\bar{u}u+\bar{d}d)/\sqrt{2}$ and
with the hidden-strange quarkonium field $\sigma_{S}\equiv \bar{s}s$ only.
Ref. \cite{dick} shows through a fit to experimental data that these
quark-antiquark states are heavy (above 1 GeV);
similarly, the scalar isovector and isodoublet states are identified with
the heavy resonances $a_{0}(1450)$ and $K_{0}^{\star }(1430),$ respectively.
Consequently, the scalar mesons below 1 GeV shall be interpreted differently:
tetraquarks or molecular/dynamically generated states are the most prominent
options, e.g. Refs. \cite{tetraquark,lowscalars} and refs. therein.

In the scalar-isoscalar sector a three-body mixing problem must be solved:
the three bare fields $\sigma _{N},$ $\sigma _{S},$ and, in addition, the
dilaton $G\equiv gg$ mix and generate the physical resonances $f_{0}(1370)$, 
$f_{0}(1500)$ and $f_{0}(1710)$. The aim is to determine the mixing matrix
in such a way that the masses and the decays of $f_{0}(1370)$, $f_{0}(1500)$
and $f_{0}(1710)$ are in agreement with the experimental results listed in
Ref. \cite{PDG}. Such a mixing problem was investigated in a variety of
phenomenological models, see Ref. \cite{review,close,scalars} and refs.
therein. However, a full calculation involving a $full$ chiral approach has
not yet been achieved for $N_{f}=3$. In Ref. \cite{dilnf3} the first step
toward this direction has been performed, but the attention was focused on
the masses only (no decays were calculated). Based on the parameter
determination obtained in\ Ref. \cite{dick}, a correct description of the
masses $f_{0}(1370)$, $f_{0}(1500)$ and $f_{0}(1710)$ implies that the
predominant glueball content is located in the resonance $f_{0}(1710)$ (and
not in $f_{0}(1500)$).

In this work we continue the study of this system by calculating, for the
first time, the decay widths of the scalar-isoscalar states. It is shown
that the mixing matrix of Ref. \cite{dilnf3} implies too large decay widths
for $f_{0}(1500)$ and $f_{0}(1710),$ and must be discarded. Therefore, we
search for other solutions. The system is extremely dependent on subtle
issues such as constructive and destructive interference effects; for this
reason, a fit (which is the most straightforward thing to do) could not yet
deliver acceptable results. Anyhow, an interesting partial solution could be
obtained by a simple `guess and try' procedure. In this solution, the gluon
condensate is very large and the glueball is to a very good extent described
by the resonance $f_{0}(1710)$ only. The phenomenology of $f_{0}(1710)$ and $%
f_{0}(1370)$ can be qualitatively described, but the kaon-kaon decay of $%
f_{0}(1500)$ is still too large. Future studies will show if this novel
solution is phenomenologically acceptable or not.

This proceeding is organized as follows: In Sec. 2 we present the effective
Lagrangian of the eLSM, in Sec. 3 we discuss the results and in Sec. 4 we
provide a summary and an outlook for work in progress.

\section{The eLSM Lagrangian}

For the purpose of studying the mixing behavior in vacuum of the
scalar-isoscalar mesons below 2 GeV we use eLSM Lagrangian developed in\
Refs. \cite{susden,dilaton, dick}:

\begin{align}
\mathcal{L}& =\mathcal{L}_{dil}+\text{\textrm{Tr}}[(D^{\mu }\Phi )^{\dag
}(D_{\mu }\Phi )]-m_{0}^{2}\left( \frac{G}{G_{0}}\right) ^{2}\text{\textrm{Tr%
}}[\Phi ^{\dag }\Phi ]-\lambda _{1}(\text{\textrm{Tr}}[\Phi ^{\dag }\Phi
])^{2}-\lambda _{2}\text{\textrm{Tr}}[(\Phi ^{\dag }\Phi )^{2}]  \notag \\
& +c_{1}(\text{det}\Phi -\text{det}\Phi ^{\dag })^{2}+\mathrm{Tr}[H(\Phi
^{\dag }+\Phi )]+\text{Tr}\left[ \left( \frac{m_{1}^{2}}{2}\left( \frac{G}{%
G_{0}}\right) ^{2}+\Delta \right) \left( L^{\mu 2}+R^{\mu 2}\right) \right] 
\notag \\
& -\frac{1}{4}\text{Tr}\left[ L^{\mu \nu 2}+R^{\mu \nu 2}\right] +\frac{h_{1}%
}{2}\text{\textrm{Tr}}[\Phi ^{\dag }\Phi ]\text{\textrm{Tr}}[L_{\mu }L^{\mu
}+R_{\mu }R^{\mu }]+h_{2}\text{\textrm{Tr}}[\Phi ^{\dag }L_{\mu }L^{\mu
}\Phi +\Phi R_{\mu }R^{\mu }\Phi ^{\dag }]  \notag \\
& +2h_{3}\text{\textrm{Tr}}[\Phi R_{\mu }\Phi ^{\dag }L^{\mu }]\text{ +... ,}
\label{Lagrangian}
\end{align}%
where $D^{\mu }\Phi =\partial ^{\mu }\Phi -ig_{1}(L^{\mu }\Phi -\Phi R^{\mu
})$. The nonstrange $\sigma _{N}\equiv \left( \bar{u}u+\bar{d}d\right) /%
\sqrt{2}$ and the strange $\sigma _{S}\equiv \bar{s}s$ bare quark-antiquark
mesons are contained in the (pseudo)scalar multiplet

\begin{equation}
\Phi =\frac{1}{\sqrt{2}}\left( 
\begin{array}{ccc}
\frac{(\sigma _{N}+a_{0}^{0})+i(\eta _{N}+\pi ^{0})}{\sqrt{2}} & 
a_{0}^{+}+i\pi ^{+} & K_{0}^{\star +}+iK^{+} \\ 
a_{0}^{-}+i\pi ^{-} & \frac{(\sigma _{N}-a_{0}^{0})+i(\eta _{N}-\pi ^{0})}{%
\sqrt{2}} & K_{0}^{\star 0}+iK^{0} \\ 
K_{0}^{\star -}+iK^{-} & \bar{K}_{0}^{\star 0}+i\bar{K}^{0} & \sigma
_{S}+i\eta _{S}%
\end{array}%
\right) .  \label{phimatex}
\end{equation}%
The matrices $L_{\mu }$ and $R_{\mu }$ describe (axial)vector fields, see
Ref. \cite{dick} for details. The scalar glueball $G\equiv gg$ is described by
the dilaton Lagrangian $\mathcal{L}_{dil}$, in which a logarithmic term with
the energy scale $\Lambda $ mimics the trace anomaly of the pure Yang-Mills
sector of QCD \cite{schechter, migdal}: 
\begin{equation}
\mathcal{L}_{dil}=\frac{1}{2}(\partial _{\mu }G)^{2}-\frac{1}{4}\frac{%
m_{G}^{2}}{\Lambda ^{2}}\left( G^{4}\ln \left\vert \frac{G}{\Lambda }%
\right\vert -\frac{G^{4}}{4}\right) \text{ .}  \label{ldil}
\end{equation}%
The three scalar fields $\sigma _{N}$, $\sigma _{S},$ and $G$ condense,
leading to the following shifts: $\sigma _{N}\rightarrow \sigma _{N}+\phi
_{N}$, $\sigma _{S}\rightarrow \sigma _{S}+\phi _{S},$ and $G\rightarrow
G+G_{0}$. As a consequence, bilinear mixing terms $\sim \sigma _{N}\sigma
_{S}$, $G\sigma _{N}$ and $G\sigma _{S}$ arise. The corresponding potential
reads:

\begin{equation}
V(\sigma _{N},G,\sigma _{S})=\frac{1}{2}(\sigma _{N},G,\sigma _{S})\left( 
\begin{array}{ccc}
m_{\sigma _{N}}^{2} & z_{G\sigma _{N}} & z_{\sigma _{S}\sigma _{N}} \\ 
z_{G\sigma _{N}} & M_{G}^{2} & z_{G\sigma _{S}} \\ 
z_{\sigma _{S}\sigma _{N}} & z_{G\sigma _{S}} & m_{\sigma _{S}}^{2}%
\end{array}%
\right) \left( 
\begin{array}{c}
\sigma _{N} \\ 
G \\ 
\sigma _{S}%
\end{array}%
\right) \text{ }  \label{pot}
\end{equation}%
where $z_{G\sigma _{N}}=-2m_{0}^{2}{\phi _{N}}/G_{0}$, $z_{G\sigma
_{S}}=-2m_{0}^{2}{\phi _{S}}/G_{0}$ and $z_{\sigma _{N}\sigma
_{S}}=2\lambda_{1}{\phi _{N}}{\phi _{S}}$. The physical states are obtained
upon diagonalization:

\begin{equation}
\left( 
\begin{array}{c}
f_{0}(1370) \\ 
f_{0}(1500) \\ 
f_{0}(1710)%
\end{array}%
\right) =B\left( 
\begin{array}{c}
\sigma _{N}\equiv \left( \bar{u}u+\bar{d}d\right) /\sqrt{2} \\ 
G\equiv gg \\ 
\sigma _{S}\equiv \bar{s}s%
\end{array}%
\right) \text{ .}  \label{trafo}
\end{equation}%
The aim is to determine the mixing matrix $B$.

\section{Results}

We use the parameters determined in\ Ref. \cite{dick}. These parameters allow
for a correct description of a variety of vacuum properties of mesons up to
1.5 GeV. However, four parameters could not be uniquely determined in Ref. \cite{dick}:
these are $m_{G} ,$ $\Lambda ,$ $\lambda _{1}$ and $h_{1}$.
In Ref. [5] three of them ($m_{G}$, $\Lambda$ and $\lambda _{1}$) were determined in order to
obtain the measured experimental masses of the resonances $f_{0}(1370)$, $%
f_{0}(1500)$ and $f_{0}(1710)$: $M_{f_{0}(1370)}=(1200$-$1500)$ MeV, $%
M_{f_{0}(1500)}=(1505\pm 6)$ MeV and $M_{f_{0}(1710)}=(1720\pm 6)$ MeV \cite%
{PDG}. A solution in which $f_{0}(1500)$ is predominantly gluonic could not
be found. On the contrary, the masses could be well described for $%
m_{G}=1.580$ GeV, $\Lambda =0.93$ GeV, and $\lambda _{1}=2.03.$ The
resulting mixing matrix reads: 
\begin{equation}
B=\left( 
\begin{array}{ccc}
0.92 & -0.39 & 0.05 \\ 
-0.22 & -0.40 & 0.89 \\ 
-0.33 & -0.83 & -0.45%
\end{array}%
\right) \text{ .}
\end{equation}%
We have now tested this scenario by evaluating the decay widths (the
corresponding mathematical expressions will be presented in a forthcoming
publication \cite{preparation}). For $h_{1}=0$ (large-$N_{c}$ limit), one
finds the decay widths into kaons and pions: $\Gamma
_{f_{0}(1710)\rightarrow \pi \pi }=0.83$ GeV, $\Gamma
_{f_{0}(1710)\rightarrow KK}=0.42$ GeV, $\Gamma _{f_{0}(1500)\rightarrow \pi
\pi }=0.22$ GeV, $\Gamma _{f_{0}(1500)\rightarrow KK}=1.14$ GeV, $\Gamma
_{f_{0}(1370)\rightarrow \pi \pi }=1.78$ GeV, $\Gamma
_{f_{0}(1370)\rightarrow KK}=0.89$ GeV. These results are clearly \emph{too
large} and cannot be cured by varying the only remaining free parameter $h_{1}$
(which should anyhow be small). Thus, the decay widths do not support this
scenario as physical. Note, such a large decay width of the predominantly
glueball state is in agreement with the study of Ref. \cite{ellis}.

The search for an acceptable solution is extremely difficult due to
interference effects in the decay amplitudes. A direct fit to the known
decay widths was so far not successful. This is why we have searched a
solutions by trying to guess the right area of the parameter space. First,
we use as an input the bare glueball mass $m_{G}=1.7$ GeV in agreement with
lattice QCD \cite{Morningstar}. Then, due to the fact that $f_{0}(1710)$ was
too broad in the previous solution, we have increased the value of the
dimensionful parameter $\Lambda $: for $\Lambda \simeq 2$ GeV the resonance $%
f_{0}(1710)$ is sufficiently narrow. By further tuning $\lambda _{1}\simeq
-10$ and $h_{1}\simeq -5,$ we obtain the mixing matrix

\begin{equation}
B=\left( 
\begin{array}{ccc}
0.90 & -0.05 & 0.41 \\ 
-0.42 & -0.03 & 0.90 \\ 
-0.04 & -0.99 & -0.05%
\end{array}%
\right) \text{ .}  \label{bnum}
\end{equation}%
The resonance $f_{0}(1710)$ is (almost) a pure glueball. The masses that are
determined by these parameters are acceptable: $M_{f_{0}(1370)}=1.06$ GeV, $%
M_{f_{0}(1500)}=1.48$ MeV and $M_{f_{0}(1710)}=1.70$ GeV. The resulting
decay widths are: $\Gamma _{f_{0}(1710)\rightarrow \pi \pi }=0.082$ GeV, $%
\Gamma _{f_{0}(1710)\rightarrow KK}=0.064$ GeV, $\Gamma
_{f_{0}(1500)\rightarrow \pi \pi }=0.14$ GeV, $\Gamma
_{f_{0}(1500)\rightarrow KK}=0.13$ GeV, $\Gamma _{f_{0}(1370)\rightarrow \pi
\pi }=0.12$ GeV, $\Gamma _{f_{0}(1370)\rightarrow KK}=0.07$ GeV.

Thus, while the decays of $f_{0}(1370)$ and $f_{0}(1710)$ are at least in
qualitative agreement with the experiment, this is not the case for $%
f_{0}(1500)$: the decays are still too large. Note also that the very large
value of $\Lambda $ implies a large gluon condensate: lattice results
\cite{Lattice} suggest that $\Lambda \lesssim 0.6$ GeV, see the discussion
in Ref. \cite{dilaton}. Thus, at this level this solution is not yet
conclusive, but can point to an interesting direction where to look for it:
large bare glueball mass in agreement with lattice (1.7 GeV) and a large
value of the gluon condensate. 

\section{Conclusions and outlook}

In this work we have studied the masses and the decays of the resonances $%
f_{0}(1370)$, $f_{0}(1500)$ and $f_{0}(1710)$ within the eLSM. Presently,
the favoured glueball seems to be $f_{0}(1710),$ in agreement with Refs. 
\cite{scalars, chen}, but further studies are needed. Namely, decay widths
which are -at least qualitatively- in agreement with data could only be
found for a large (eventually too large!) value of the gluon condensate. 

Another possibility is an improvement of the underlying effective model of
Ref. \cite{dick}, by studying the influence of a quadratic mass term in the
(pseudo)scalar sector. This is a minimal change of Ref. \cite{dick}, which
however can have interesting phenomenological implications due to the fact
that the strange current quark mass is not negligible. For value of the
gluon condensate in agreement with lattice, a not too broad glueball can
only be found if destructive interferences between the different amplitudes
occur. This is why an improved numerical analysis, which allows to study in
detail the whole parameter space, would be also helpful.

More in general, one can extend the study of glueballs with other quantum numbers. In
Ref. \cite{psg} the pseudoscalar glueball was investigated within the eLSM.
A similar program can be carried out for a tensor glueball with a mass of
about 2.2 GeV, see e.g. Ref. \cite{tensor} and refs. therein, as well as
heavier glueballs, such as the vector and pseudovector glueball states, with
lattice predicted masses of about $3.8$ and $3$ GeV, respectively.

\section*{Acknowledgments}

The authors thank I. Mishustin, D. Parganlija, and D. H. Rischke for useful
discussions. S. J. acknowledges support from H-QM and HGS-HIRe.

\end{document}